# Localization Principle of the Spectral Expansions of Distributions Connected with Schodinger Operator

[1]Abdumalik Rakhimov, [2]Anvarjon Ahmedov and [3]Hishamuddin Zainuddin

[1]Department of Mathematics, National University of Uzbekistan, Tashkent, Uzbekistan
[2]Department of Process and Food Engineering, Faculty of Engineering, Universiti Putra Malaysia, Serdang, Selangor, Malaysia
[3]Institute for Mathematical Research, Universiti Putra Malaysia, Serdang, Selangor, Malaysia

**Abstract:** In this paper the localization properties of the spectral expansions of distributions related to the self adjoint extension of the Schrodinger operator are investigated. Spectral decompositions of the distributions and some classes of distributions are defined. Estimations for Riesz means of the spectral decompositions of the distributions in the norm of the Sobolev classes with negative order are obtained.

## INTRODUCTION

Many phenomena in nature require for its description either "bad" functions or even they can not be described by regular functions. As an example one can consider Dirac's delta function which occurs in many problems of Quant Mechanics and Modern Mathematical Physics. Singularity of this function creates some mathematical problems in applications of methods of solutions (for instant Fourier method).

As a rule in such situations the solutions can be approximated by solutions of the regularized problems and hence it is important to study convergence and summability corresponding series and/or integrals (spectral expansions). Nevertheless, distributions, as well as delta function of Dirac, describe only integral characteristics of phenomena. Hence so it is necessary to prove applicability developed methods in the spaces of distributions.

For instant one can study these problems in domains where a distribution is equal zero. Such a problem, following Riemann, called localization problems. Localization of spectral expansions of distribution (convergence to zero in domains where distribution coincides with zero) is very difficult problem due to influence of singularities. One of the classical examples for localization problem is the fact that the partial sums of Fourier series of Dirac's delta function is divergent, the latter coincides with Direchlet's kernel of Fourier series. But regularized kernel (Fejer's kernel)is convergent (in one dimensional case). And multidimensional case requires more regularization power.

For the first time sharp conditions for regularization of spectral expansions in Sobolev's spaces where established by Sh.A. Alimov (1993). Further this theory where developed for the more general spectral expansions by other authors. In particular this problem was considered for the operators with singular coefficients. In (Rakhimov, 2003) localization problem for eigenfunction expansions connected with Schrödinger's operator in bounded domain in the spaces of distributions considered and sharp conditions were established.

In the present work we consider spectral expansions connected with Schrödinger's operator in whole space, and extend the localization properties of the spectral expansions related to distributions.

*Preliminaries:*

We consider the Schrodinger operator $L = -\Delta + q(x)$ with domain $C_0^\infty(R^N)$, where $q(x)$ is potential with singularity at 0 satisfies following conditions

$$\frac{\partial^{|\alpha|} q(x)}{\partial^{\alpha_1} x_1 \partial^{\alpha_2} x_2 ..... \partial^{\alpha_N} x_N} | \leq const |x|^{-1-\alpha},$$

**Corresponding Author:** Anvarjon Ahmedov, Department of Process and Food Engineering, Faculty of Engineering, Universiti Putra Malaysia, Serdang, Selangor, Malaysia
E-mail: anvar@eng.upm.edu.my

where $\alpha = (\alpha_1, \alpha_2, ......, \alpha_N)$ is multi index-vector with integer non negative coordinates and $|\alpha| = \sum_{j=1}^{N} \alpha_j$.

From Kato-Rellich theorem (Reed and Simon, 1980) it follows that operator L is essentially self-adjoint and semibounded. A lower bound of Schrodinger operator it is denoted by $\mu$. The self-adjoint extension $\hat{L}$ in $L_2(R^N)$ of the Schrodinger operator coincides with its closure. Let $\{E_\lambda\}$ be the spectral decomposition of unity corresponding to $\hat{L}$. It is well known that the operators $E_\lambda$ are integral operators whose kernels $\Theta(x, y, \lambda)$ belong to the space $C^\infty(R^N)$ with respect to the both variables $x$ and $y$ and for any $\lambda$.

The spectral decomposition of an arbitrary function $g \in L_2(R^N)$ can be defined by

$$E_\lambda g(x) = \int_{R^N} g(y) \Theta(x, y, \lambda) dy.$$

By $\varepsilon'(R^N)$ we denote the space of linear continuous functionals on $C^\infty(R^N)$. Since $\Theta(x, y, \lambda) \in C^\infty(R^N \times R^N)$, it follows that for any $f \in \varepsilon'(R^N)$ we can define the spectral decomposition of $f$ by the formula

$$E_\lambda f(x) = \langle f, \Theta(x, y, \lambda) \rangle,$$

where the functional $f$ acts on $\Theta(x, y, \lambda)$ with respect to the second argument. While the Riesz means of the spectral decomposition of $f$ can be defined as follows

$$E_\lambda^s f(x) = \langle f, \Theta^s(x, y, \lambda) \rangle, s \geq 0$$

where

$$\Theta^s(x, y, \lambda) = \int_\mu^\lambda \left( \frac{\lambda - t}{\lambda - \mu} \right)^s d_t \Theta(x, y, t).$$

Note that for any $f \in \varepsilon'(R^N)$ its spectral expansions $E_\lambda^s f(x)$ belongs to the space $C^\infty(R^N)$ for any $\lambda > 0$.

For any real $\ell$, by $H^\ell(R^N)$ we denote the Sobolev classes (Egorov, 1984). When $\ell > 0$ there are Sobolev classes $W_2^\ell$ and for negative values of $W_2^\ell$ conjugates for positive classes.

## RESULTS AND DISCUSSIONS

In the present paper we establish the sharp conditions for localization of Riesz means of spectral expansions connected with Schrödinger's operator in whole space. These conditions describe relationship between singularities of distributions from negative Sobolev's spaces, power of the regularization and dimension of the space.

The main results of the paper are the following

***Theorem 3.1.***

Let $\ell > 0, s \geq 0,$ and $f \in \varepsilon'(R^N) \cap W_2^{-l}(R^N)$. If $s \geq (N-l)/2 + \ell$, then

$$\lim_{\lambda \to \infty} E_\lambda^s f(x) = 0$$

uniformly with respect to $x \in K$ for any compact subset $K \subset R^N \, supp(f)$.

***Corollary 3.2:***

Let $f \in \varepsilon'(R^N) \cap W_2^{-l}(R^N)$, $l > 0$ and let the distribution $f$ coincide with a continuous function $g(x)$ in a domain $D$. If $s \geq (N-l)/2 + \ell$, then

$$\lim_{\lambda \to \infty} E_\lambda^s f(x) = 0$$

uniformly on each compact set $K \subset D$.

The following are key lemmas in the proof of the main theorem.

***Lemma 3.3:***

Let $\Omega$ an arbitrary bounded domain in $R^N$ and $s \geq 0$. Then for any function $g \in L_2(R^N)$, with $supp \, g \subset \Omega$ and any compact set $K \subset R^N \setminus \overline{\Omega}$ uniformly by $x \in K$

$$|E_\lambda^s g(x)| \leq c(K) \|g\|_{L_2} \lambda^{\frac{N-1-2s}{4}}$$

***Lemma 3.4:***

Let $\Omega$ an arbitrary bounded domain in $R^N$ and $s \geq 0$ Then for any compact set $K \subset R^N \setminus \overline{\Omega}$ uniformly by $x \in K$

$$\|\Theta^s(x, y, \lambda)\|_{L_2(\Omega)} \leq c(K) \lambda^{\frac{N-1-2s}{4}}$$

Using von Niemann's spectral theorem for selfadjoint operators in Hilbert spaces for arbitrary number $\tau \geq 0$ define powers of the operator $\hat{L}^\tau f(x)$ by

$$\hat{L}^\tau f(x) = \int_0^\infty t^\tau dE_t f(x).$$

Then define a function

$$\Theta_\tau^s(x, y, \lambda) = \hat{L}^\tau \Theta^s(x, y, \lambda)$$

which by can be represented as follows

$$\Theta_\tau^s(x, y, \lambda) = \int_0^\lambda t^\tau d_t \Theta^s(x, y, t).$$

Using integration by part we obtain the following equality $\Theta_\tau^s(x, y, \lambda) = \lambda^\tau \Theta^s(x, y, \lambda) - \tau \int_0^\lambda t^{\tau-1} \Theta^s(x, y, t) dt$. This equality together with Lemma 3.4 gives

***Lemma 3.5:***

Let $\Omega$ an arbitrary bounded domain in $R^N$ and $s \geq 0$, $\tau > 0$. Then for any compact set $K \subset R^N \setminus \overline{\Omega}$ uniformly by $x \in K$

$$\|\Theta_\tau^s(x, y, \lambda)\|_{L_2(\Omega)} \leq c(K) \lambda^{\frac{N-1-2s}{4} + \tau}$$

Let $f \in \varepsilon'(R^N) \cap H^{-\ell}(R^N)$ and $\ell > 0$. Since any distribution from $\varepsilon'(R^N)$ is compactly supported in $R^N$, it follows that there exists a bounded domain $\Omega \subset R^N$ such that the inequality

$$|\langle f, u \rangle| \leq \|f\|_{-\ell} \|u\|_{W_2^\ell(\Omega)}$$

is valid for all $u(x) \in C^\infty(\Omega)$, and $\|\cdot\|_{-\ell}$ means norm in the space $H^{-\ell}$. Domain $\Omega$ can be chosen from the condition $supp\ f \subset \Omega$. Then taking $u(y) = \Theta^s(x, y, \lambda)$ from last inequality we will obtain

$$|E_\lambda^s f(x)| \leq \|f\|_{-\ell} \|\Theta^s(x, y, \lambda)\|_{W_2^\ell(\Omega)}$$

Then taking $\tau = \dfrac{\ell}{2}$ from Lemma 3.3 one can obtain inequality

$$|E_\lambda^s f(x)| \leq c(K) \sqrt{\lambda}^{\frac{N-1}{2} + \ell - s} \|f\|_{-\ell}.$$

From this inequality statement of the Theorem 3.1 follows immediately. Corollary 3.2 follows from Theorem 3.1 and the fact that for continuous function with compact support uniformly convergence of Reisz means of spectral expansions valid as soon as $s > \dfrac{N-1}{2}$.

Note that in (Rakhimov, 1996) for Fourier Integral with spherical partial sums (which corresponds for the case $q(x) = 0$) it was proved sharpness of the condition $s \geq (N-l)/2 + \ell$. Thus conditions of the Theorem 3.1 are sharp.

## ACKNOWLEDGMENT

This work was done during the first authors visit to the Institute for Mathematical Research (INSPEM), Universiti Putra Malaysia. The first author gratefully acknowledges INSPEM for support and hospitality. The second and third authors were supported by Universiti Putra Malaysia under Research University Grant (RUGS), project number 05-01-09-0674RU and the Institute for Mathematical Research Universiti Putra Malaysia under Science Fund Grant Scheme (SF)(project number is 06-01-04-SF0256) and under Fundamental Research Grant Scheme (FRGS) (project code 05-10-07-379 FR).